\documentclass[twocolumn,pra,floatfix,amsmath,superscriptaddress]{revtex4-1}
\usepackage[latin9]{inputenc}
\usepackage{graphicx}
\usepackage[hypertex]{hyperref}
\usepackage{bm,bbm}
\usepackage{amsmath}
\usepackage{amsfonts}
\usepackage{amsthm}
\usepackage{amssymb}
\usepackage{mathrsfs}
\usepackage{subfigure}
\usepackage{array}
\usepackage{color}
\makeatletter

\usepackage{pxfonts}

\newtheorem{theorem}{Theorem}
\newtheorem{lemma}{Lemma}

\begin{document}

\title{Maximal coherence in generic basis}

\author{Yao Yao}
\email{yaoyao@mtrc.ac.cn}
\affiliation{Microsystems and Terahertz Research Center, China Academy of Engineering Physics, Chengdu Sichuan 610200, China}
\affiliation{Institute of Electronic Engineering, China Academy of Engineering Physics, Mianyang Sichuan 621999, China}

\author{G. H. Dong}
\affiliation{Beijing Computational Science Research Center, Beijing, 100094, China}

\author{Li Ge}
\affiliation{School of Science, Hangzhou Dianzi University, Hangzhou Zhejiang 310018, China}

\author{Mo Li}
\email{limo@mtrc.ac.cn}
\affiliation{Microsystems and Terahertz Research Center, China Academy of Engineering Physics, Chengdu Sichuan 610200, China}
\affiliation{Institute of Electronic Engineering, China Academy of Engineering Physics, Mianyang Sichuan 621999, China}

\author{C. P. Sun}
\affiliation{Beijing Computational Science Research Center, Beijing, 100094, China}
\affiliation{Synergetic Innovation Center of Quantum Information and Quantum Physics, University of Science and Technology of China,
Hefei, Anhui 230026, China}

\date{\today}

\begin{abstract}
Since quantum coherence is an undoubted characteristic trait of quantum physics, the quantification and application of
quantum coherence has been one of the long-standing central topics in quantum information science. Within the framework of
a resource theory of quantum coherence proposed recently, a \textit{fiducial basis} should be pre-selected for characterizing
the quantum coherence in specific circumstances, namely, the quantum coherence is a \textit{basis-dependent} quantity.
Therefore, a natural question is raised: what are the maximum and minimum coherences contained in a certain quantum state with respect to a generic basis?
While the minimum case is trivial, it is not so intuitive to verify in which basis the quantum coherence is maximal. Based on the coherence measure of
relative entropy, we indicate the particular basis in which the quantum coherence is maximal for a given state,
where the Fourier matrix (or more generally, \textit{complex Hadamard matrices}) plays a critical role
in determining the basis. Intriguingly, though we can prove that the basis associated with the Fourier matrix is a
stationary point for optimizing the $l_1$ norm of coherence, numerical simulation shows that it is not a global optimal choice.
\end{abstract}


\maketitle
\section{INTRODUCTION}
Quantum coherence, as a prominent resource for quantum information processing, has found its various
diversified applications in quantum cryptography \cite{Gisin2002}, quantum computation \cite{Cleve1997,Nielsen}
and quantum metrology \cite{Giovannetti2004,Giovannetti2006}. Nevertheless, until very recently, there
still lacks a rigorous information-theoretic framework for characterizing coherence and a great deal of effort
has been devoted to this significant and long-standing topic \cite{Streltsov2016}. Exploiting the quantum resource theory \cite{Horodecki2013,Brandao2015},
Baumgratz \textit{et al.} proposed a framework for quantifying coherence based on distance or pseudo-distance measures \cite{Baumgratz2014},
by noting that a similar line of thought has been successfully applied in the theory of quantum entanglement \cite{Vedral1997,Vedral1998}.
Within this framework, several reasonable postulates has been proposed which should be satisfied by all \textit{bona fide} measures of quantum coherence.
Moreover, as a prerequisite, a \textit{fiducial basis} should be pre-selected for determining the exact value of quantum coherence,
according to the specific theoretic considerations or physical implementations. In other words, the coherence measures
defined in Ref. \cite{Baumgratz2014} are all \textit{basis-dependent} quantities. For a simple example, the eigenvectors of Pauli
matrices $\sigma_x$ and $\sigma_z$ constitute two mutually unbiased bases, that is, each incoherent basis pure state
is a maximally coherent state with respect to the other basis \cite{Durt2010}.

Therefore, a natural question is raised: what are the maximum and minimum coherences contained in a given quantum state
with respect to a generic basis? Notably, this problem is not only theoretically motivated but also experimentally
relevant. Since quantum coherence has been identified as the essential resources for certain quantum information tasks,
it will be preferred to extract the coherence content of a given state as much as possible. Obviously, the key issue is to
find out a reference basis with respect to which the coherence value is maximal. It is noteworthy that a similar but distinct
problem has been discussed by Singh \textit{et al.} \cite{Singh2015}, where a notion of maximally coherent mixed states
(MCMSs) was considered and in fact they obtained the maximally achievable quantum coherence for a fixed mixedness in the
computational basis. However, while in Ref. \cite{Singh2015} the purity is the only independent variable, here we
expect that the maximally achievable value of quantum coherence may depend on the entropy or eigenvalues of the given state.

In this work, we mainly focus on the coherence measures of relative entropy and $l_1$ norm, which are the only monotones
that are found to satisfy all criterions proposed in \cite{Baumgratz2014} until now. While any pure state can always be
transformed to a maximally coherent state by a change of basis, we realize that for general mixed states
the situation becomes much more complicated and subtle. For the relative entropy of coherence, we demonstrate that
the basis associated with the Fourier matrix (in fact, all complex Hadamard matrices)
is optimal for achieving the maximal coherence. Since all \textit{bona fide} measures of quantum coherence
satisfy the same set of constraints, intuitively one might be tempted to conjecture that
this particular basis is also optimal for other measures of quantum coherence.
However, although we prove that this particular basis is a stationary point (e.g., a local extremum) for optimizing the $l_1$ norm of coherence, the
numerical simulation shows that in general it is not a global optimal choice, especially
for high-dimensional mixed states. Therefore, this seemingly counter-intuitive finding illustrates that
the condition for achieving maximum values of coherence is not universal, but measure-dependent.

The paper is organized as follows. In Sec. \ref{sec2}, we briefly
review the resource framework of quantum coherence and define the problem in standard notations. In
Sec. \ref{sec3}, we identify the basis in which the relative entropy of coherence of a given state is maximal
and the significance of the Fourier matrix (complex Hadamard matrices) is illustrated.
In Sec. \ref{sec4}, we prove that this basis is a stationary point for optimizing the $l_1$ norm of coherence
and the properties of the circulant matrix is emphasized.
In Sec. \ref{sec5}, we perform a detailed numerical simulation and demonstrate that the basis associated with
the Fourier matrix is not global optimal.
The Sec. \ref{sec6} is devoted to the conclusion and discussion
of the main results and several open questions are presented for future investigation.

\section{Defining the problem}\label{sec2}
Throughout the paper, we adopt the resource theory of quantum coherence proposed
by Baumgratz \textit{et al.} in \cite{Baumgratz2014}. A general framework of quantum resource theory consists of three
key ingredients: (i) the free states, (ii) the resource states, and (iii) the
restricted or free operations \cite{Horodecki2013,Brandao2015}.
Quantum entanglement theory is another prominent and familiar application of this theoretical framework,
where the three basic ingredients are separable states, entangled states and LOCC, respectively \cite{Vedral1997,Vedral1998}.
Except for these basic notions, such as the free states and free operations defined in their own context,
the quantum resource theories mainly rely on the following two aspects: (i) a series of reasonable postulates
that should be fulfilled by each measure or indicator of genuine resource; (ii) a set of contractive
geometric metrics \cite{Brandao2015}. In the corresponding resource theory of coherence, the free
(incoherent) states are those diagonal in a pre-selected basis in the $d$-dimensional Hilbert space $\mathcal{H}$,
denoted by the set $\mathcal{I}$. Accordingly, the free (incoherent) operations are completely positive and trace preserving
quantum maps $\Phi_{\mathcal{I}}$ admitting an operator-sum representation where every Kraus operator $K_i$ will transform
the set of incoherent states into itself \cite{Baumgratz2014}, that is, $K_i\mathcal{I}K_i^\dagger\subset\mathcal{I}$.

Except for the nullity condition and convexity requirement, each \textit{bona fide} measures of quantum coherence
$C(\rho)$ are assumed to be a monotone function under the non-selective and sub-selective incoherent measurements, namely \cite{Baumgratz2014}
\begin{align}
C(\Phi_{\mathcal{I}}(\rho))\leq C(\rho), \\
\sum_ip_iC(\rho_i)\leq C(\rho),\label{SM}
\end{align}
where $p_i=\textrm{tr}(K_i\rho K_i^\dagger)$ and $\rho_i=K_i\rho K_i^\dagger/p_i$. Note that the latter constraint
combined with convexity condition will lead to the former, which implies the latter is a stronger monotonicity requirement.
Based on these criteria, several potential candidates were put forward for the quantification of coherence \cite{Baumgratz2014}.
However, so far only two measures have been identified to fulfill all the requirements, that is, the relative entropy of coherence $C_\mathcal{R}(\rho)$
and the $l_1$ norm of coherence $C_{l_1}(\rho)$
\begin{align}
C_\mathcal{R}(\rho)&=S(\rho_{\mathcal{I}})-S(\rho), \\
C_{l_1}(\rho)&=\sum_{\mu\neq\nu}|\rho_{\mu\nu}|,
\end{align}
where $\rho_{\mathcal{I}}$ is the diagonal part of $\rho=\sum_{\mu,\nu}\rho_{\mu\nu}|\mu\rangle\langle \nu|$. It is worth pointing out that the coherence measures induced by $l_2$ norm
and the fidelity do not constitute valid coherence monotones, since for both quantities the strong monotonicity (\ref{SM}) does not hold
in general \cite{Baumgratz2014,Shao2015}. Moreover, though the trace-norm measure of coherence was proved to be a strong monotone for all qubit and $\textrm{X}$ states,
a recent work showed that the trace norm of coherence cannot be regarded as a legitimate coherence measure for general states \cite{Yu2016}.

In such a framework, a pre-determined \textit{fiducial basis} is prior to any evaluation of the value of coherence.
Recall that a specific orthonormal basis corresponds to a particular matrix representation of a given density matrix
(the maximally mixed state $\rho_\star=\openone/d$ is an exception since it is always diagonal in any bases).
From the definitions of $C_\mathcal{R}(\rho)$ and $C_{l_1}(\rho)$, it is easy to see that any density matrix
with off-diagonal entries in such a representation will be identified as a resource state (having non-zero coherence value)
with respect to this fiducial basis. Furthermore, it is noteworthy that any two distinct orthonormal bases are connected with
a unitary operator \cite{Nielsen} and this remarkable fact indicates that with reference to the computational basis $\{|i\rangle\}_{i=1}^d$,
any generic basis $\{|a_i\rangle\}_{i=1}^d$ can be fully characterized by a unitary operator $U$,
with a \textit{one-to-one correspondence} $|a_i\rangle=U|i\rangle$. Therefore, for a given density matrix $\rho$, the evaluation of quantum coherence
in a generic basis $\{|a_i\rangle\}$ is equivalent to considering the coherence of $U^\dagger\rho U$ in the the computational basis
$\{|i\rangle\}$, that is
\begin{align}
\langle a_i|\rho|a_j\rangle=\langle i|U^\dagger\rho U|j\rangle, \quad i,j=1,\ldots,d.
\label{ER}
\end{align}

Therefore, the problem of determining in which basis the matrix representation will display the maximum or minimum coherence
is tantamount to finding out the corresponding unitary transformations. In fact, according to the spectral theorem \cite{Nielsen,Watrous},
any normal operator is diagonalizable and thus, for a given density matrix $\rho$, the minimum coherence is always zero
in which case the eigenvectors of $\rho$ form the columns of this particular $U$. However, on the other hand, it is not immediately intuitive that
in which basis we can acquire the maximum coherence.

\section{Relative entropy of coherence}\label{sec3}
\subsection{Complex Hadamard matrices}

Now let us consider a fixed density matrix $\rho$ in the $d$-dimensional Hilbert space $\mathcal{H}$.
In this section we will mainly concentrate on $C_\mathcal{R}$, since the entropy function solely depends
on the eigenvalues of its argument and is usually easier to handle. To begin with, two observations
caught our attention. First, regardless of the fiducial basis, $C_\mathcal{R}$ is universally upper bounded by
\begin{align}
0\leq C_\mathcal{R}(\rho)&=S(\rho_{\mathcal{I}})-S(\rho)\leq \log d-S(\rho).
\label{UB}
\end{align}
In fact, this inequality stems from the majorization relation $\textrm{diag}\{1/d,\ldots,1/d\}\prec \rho_{\mathcal{I}}\prec \rho$,
also known as the Schur-Horn theorem \cite{Bengtsson}.
Note that though $\rho$ itself always remain unchanged with respect to the basis change, the diagonal part $\rho_{\mathcal{I}}$
will definitely depend on different matrix representations. The tightness of this upper bound is equivalent to
whether there exists a specific basis in which the matrix representation of $\rho$ has equal main diagonal elements.
The second observation is recapitulated in the following lemma, which was proved by Horn and Johnson using two different approaches
\cite{Horn1985,Horn1991}.

\begin{lemma}
Denote the set of $d$-dimensional square matrices by $\mathcal{M}_d$.
Then for each $A\in \mathcal{M}_d$, there exists a unitary matrix $U\in \mathcal{M}_d$ such
that all the diagonal entries of $U^\dagger AU$ have the same value $\textrm{tr}A/d$.
\end{lemma}

This lemma is a rather general result and implies that $C_\mathcal{R}$ can always achieve the upper bound of Eq. (\ref{UB})
through the change of basis. However, the proofs in Ref. \cite{Horn1985,Horn1991} do not indicate the explicit
form of this particular $U$. The following theorem spells out the exact form of this type of $U$ and thus the corresponding basis,
where the complex Hadamard matrices play an essential role. Recall that a complex Hadamard matrix $H$
is commonly defined as a $d$-dimensional square matrix with the properties of \textit{unimodular} and \textit{orthogonality}
\cite{Tadej2006,Bengtsson2007,Szollosi2011}
\begin{align}
|H_{ij}|&=1, \forall \, i,j=0,\ldots,d-1, \\
H H^\dagger&=d\openone.
\end{align}

\begin{theorem}
There exists a set of unitary operators $U$ such that $C_\mathcal{R}$ achieves the maximum value $\log d-S(\rho)$
for a given density matrix $\rho$, where the reference basis is defined by $\{U|i\rangle\}_{i=0}^{d-1}$.
The unitary transformation has the form $U=V H^\dagger$, where $V$ consists of the eigenvectors of $\rho$
as its column vectors and $H$ belongs to the set of (rescaled) complex Hadamard matrices.
\label{T1}
\end{theorem}

\textit{Proof.} From Eq. (\ref{ER}), the evaluation of coherence value of $\rho$ in the transformed basis
$\{U|i\rangle\}_{i=0}^{d-1}$ is equivalent to that of $U^\dagger \rho U$ in the computational basis.
Due to the spectral decomposition of $\rho=V\Lambda V^\dagger$, without any loss of generality,
we assume
\begin{align}
\Lambda=\textrm{diag}\{\lambda_0,\lambda_1,\ldots,\lambda_{d-1}\}, \quad \lambda_i\geq\lambda_{i+1},
\end{align}
by proper arrangement of the order of the eigenvectors in $V$. Then we have
\begin{align}
U^\dagger \rho U=H V^\dagger\rho V H^\dagger=H \Lambda H^\dagger.
\end{align}
Let us denote the elements of the matrix $A$ as $[A]_{ij}=A_{ij}$. Adopting the Einstein convention, the diagonal
entries in this matrix representation are
\begin{align}
[H \Lambda H^\dagger]_{ii}&=[H]_{ik}\lambda_k\delta_{kl}[H^\dagger]_{li} =\lambda_kH_{ik}H^\ast_{ik} \nonumber\\
&=\frac{1}{d}\sum_k\lambda_k=\frac{1}{d},
\end{align}
where we prefer the \textit{rescaled} definition of complex Hadamard matrices, that is, $H H^\dagger=\openone$
with complex entries of equal modulus $|H_{ij}|=1/\sqrt{d}$.  \hfill $\blacksquare$

To gain a better understanding of how the theorem works, we would like to take some time to further illustrate the notion
of complex Hadamard matrices.
Here we rescale a complex Hadamard matrix to a corresponding unitary matrix, and thus the $d$ vectors formed by the columns
of such a matrix constitute a complete set of orthogonal basis of $\mathbb{C}^d$. It is noteworthy that
each of this set of basis vectors is mutually unbiased with respect to the computational basis. That is,
following the notations of Ref. \cite{Baumgratz2014}, these basis vectors are \textit{maximally coherent states} in the standard basis.
Another important concept is the equivalence relation between two different complex Hadamard matrices.
Two Hadamard matrices $H_1$ and $H_2$ are called equivalent, denoted by $H_1\simeq H_2$,
if there exist diagonal unitary matrices $D_1$ and $D_2$ and permutation matrices $P_1$ and $P_2$ such that \cite{Tadej2006,Bengtsson2007,Szollosi2011}
\begin{align}
H_1=D_1P_1H_2P_2D_2=M_1H_2M_2,
\end{align}
where $M_1=D_1P_1$ and $M_1=P_2D_2$ are so-called generalized permutation matrices or monomial states \cite{Nest2011},
which are unitary matrices with the matrix representation in the standard basis containing
precisely one nonzero entry in each row and column.

Therefore, the reordering of the rows and columns or rephasing of the off-diagonal entries of a complex Hadamard matrix
does not alter its equivalence class. Consequently, every complex Hadamard matrix can be transformed to a \textit{dephased} form,
where the entries of its first row and column are all equal to $1/\sqrt{d}$ \cite{Tadej2006,Bengtsson2007,Szollosi2011}.
An important example is the Fourier matrix, which exists for all dimensions and is naturally of the dephased form
\begin{align}
[\mathbf{F}_d]_{\mu\nu}=\frac{1}{\sqrt{d}}\textrm{e}^{\frac{2\pi \textrm{i}\mu\nu}{d}}=\frac{1}{\sqrt{d}}\omega^{\mu\nu}, \quad \mu,\nu=0,\ldots,d-1
\label{Fourier}
\end{align}
where we denote by $\omega=\textrm{e}^{2\pi\textrm{i}/d}$ the $d$-\textrm{th} root of unity. On the other hand,
if we choose the standard basis $\{|i\rangle\}_{i=0}^{d-1}$ to be incoherence basis, then the diagonal unitary matrices $D$
and permutation matrices $P$ (and thus monomial matrices $M$) in such a basis are all incoherent unitary operators, due to
the criterion derived in Ref. \cite{Yao2015}. Moreover, since the inverse matrices of $D$ and $P$ are still incoherent operations, all these incoherent
unitary matrices are coherence-value-preserving operations (CVPOs) \cite{Peng2016}. Intriguingly, it is easy to prove that
the CVPOs admitted by all valid coherence measures are monomial matrices, which in fact are combinations of rephasing and relabeling.
Therefore, due to its simplicity and significance, henceforth we adopt the Fourier matrix as the primary representative
of complex Hadamard matrices, though some of the following conclusions also hold for this whole set of matrices.

\subsection{Dual basis}
For a qubit system, the Fourier matrix is just the so-called Hadamard gate of quantum computation
\begin{align}
\mathbf{F}_2=\frac{1}{\sqrt{2}}
\left(\begin{array}{cc}
1 & 1 \\
1 & \omega
\end{array}\right)=\frac{1}{\sqrt{2}}
\left(\begin{array}{cc}
1 & 1 \\
1 & -1
\end{array}\right).
\end{align}
By applying the Hadamard gate to the standard basis $\{|0\rangle,|1\rangle\}$,
the following two vectors can also be obtained
\begin{align}
|\phi_\mu\rangle=\mathbf{F}_2|\mu\rangle=\frac{1}{\sqrt{2}}\left[|0\rangle+(-1)^\mu|1\rangle\right],\quad \mu=0,1
\end{align}
Obviously, the transformed basis $\{|\phi_0\rangle,|\phi_1\rangle\}$ is mutually unbiased with respect to $\{|0\rangle,|1\rangle\}$,
or equivalently, each basis state in one set is the maximally coherent state with respect to another. In this context, the Hadamard gate
can be termed as a \textit{maximally coherent operator} for a qubit system \cite{Yao2015}. As a generalization of the Hadamard gate
in arbitrary finite dimension, the Fourier matrix (in fact, all complex Hadamard matrices) inherits the properties of the Hadamard gate.
Namely, we can obtain a so-called \textit{dual basis} by applying $\mathbf{F}_d$ on the computational basis
\begin{align}
|\phi_k\rangle=\mathbf{F}_d|k\rangle=\frac{1}{\sqrt{d}}\sum_{n=0}^{d-1}\omega^{k n}|n\rangle,\quad k=0,1,\ldots,d-1
\end{align}
To certify the orthogonality of the basis states, one can verify the overlap
\begin{align}
\langle \phi_\nu|\phi_\mu\rangle=\frac{1}{d}\sum_{n=0}^{d-1}\omega^{(\mu-\nu)n}=\delta_{\mu,\nu}
\label{orthogonality}
\end{align}
where $\delta_{\mu,\nu}$ is the Kronecker delta function. Remarkably, $\{|j\rangle\}$ and $\{|\phi_j\rangle\}$ are the eigenvectors of
the generalized Pauli operator $\mathbb{Z}_d$ and $\mathbb{X}_d$, respectively \cite{Gottesman2001,Bandyopadhyay2002}
\begin{align}
\mathbb{Z}_d|j\rangle=\omega^j|j\rangle,\quad \mathbb{X}_d|j\rangle=|j+1\rangle,
\end{align}
and further we have
\begin{align}
\mathbb{Z}_d|\phi_j\rangle=|\phi_{j+1}\rangle,\quad \mathbb{X}_d|\phi_j\rangle=\omega^{-j}|\phi_j\rangle.
\end{align}

In fact, in terms of $\{|\phi_j\rangle\}$ and the eigenvectors $\{|\psi_j\rangle\}$ of $\rho=\sum_j\lambda_j|\psi_j\rangle\langle\psi_j|$,
the unitary operators $V$ and $\mathbf{F}_d$ defined in Eq. (\ref{Fourier}) can be written as
\begin{gather}
V=\sum_{j=0}^{d-1}|\psi_j\rangle\langle j|, \\
\mathbf{F}_d=\frac{1}{\sqrt{d}}\sum_{i,j=0}^{d-1}\omega^{ij}|i\rangle\langle j|=\sum_{j=0}^{d-1}|\phi_j\rangle\langle j|.
\end{gather}
Therefore, $U=V\mathbf{F}_d^\dagger=\sum_j|\psi_j\rangle\langle\phi_j|$ and the transformed state $U^\dagger\rho U$ is given by
\begin{align}
U^\dagger\rho U=\sum_{j=0}^{d-1}\lambda_j|\phi_j\rangle\langle\phi_j|.
\end{align}
Since $|\phi_j\rangle$ are maximally coherent states in the computational basis, the transformed state $U^\dagger\rho U$ is a
weighted mixture of $\{|\phi_j\rangle\}$ and thus has equal main diagonal entries with respect to the standard basis.

\subsection{$l_2$ norm of coherence}
As a byproduct, we can arrive at a conclusion that $l_2$ norm of coherence $C_{l_2}$ also achieves the maximum value exactly in the same basis,
though the strong monotonicity condition is not satisfied by $C_{l_2}$. First, for a given density matrix
$\rho=\sum_{\mu,\nu}\rho_{\mu\nu}|\mu\rangle\langle\nu|$, the $l_2$ norm of coherence
is defined by \cite{Baumgratz2014}
\begin{align}
C_{l_2}(\rho)=\sum_{\mu\neq\nu}|\rho_{\mu\nu}|^2=\|\rho\|_2^2-\sum_{\mu=0}^{d-1}|\rho_{\mu\mu}|^2,
\label{L2}
\end{align}
where $\|\bullet\|_2$ is called the Hilbert-Schmidt norm or the Frobenius norm (and is occasionally written as $\|\bullet\|_\textrm{F}$
for that reason) \cite{Horn1985,Bhatia}. An important property of the Frobenius norm is the unitary invariance, that is, for any
$A\in\mathcal{M}_d$ and arbitrary unitary matrices $U,V\in\mathcal{M}_d$
\begin{align}
\|UAV\|_2^2=\|A\|_2^2=\textrm{tr}(A^\dagger A)=\sum_{\mu,\nu}|A_{\mu\nu}|^2.
\end{align}
Therefore, basis change does not alter the Frobenius norm of a given density matrix and one can only focus on
the diagonal parts of matrix representations for distinct bases, as can be seen from Eq. (\ref{L2}).

Moreover, in an arbitrary basis, the diagonal part of the corresponding matrix representation
constitutes a nonnegative vector of $\mathbb{R}^d$ with elements summing to unity.
Denote such a vector by $\mathbf{e}_{\mathcal{I}}$ and the uniformly distributed probability vector
by $\mathbf{e}_0$. According to the majorization theory, $\mathbf{e}_0$ is majored by
arbitrary vector $\textrm{e}_{\mathcal{I}}$, that is \cite{Bengtsson,Bhatia}
\begin{align}
\mathbf{e}_0=\frac{1}{d}(1,1\ldots,1)\prec\mathbf{e}_{\mathcal{I}}=(\{\rho_{\mu\mu}\}_{\mu=0}^{d-1}),
\end{align}
where $\rho_{\mu\mu}=\langle\mu|\rho|\mu\rangle$ and $\{|\mu\rangle\}$ is the specified incoherent basis
(not necessarily the computational basis). Since the function $f(\mathbf{x})=\sum_ix_i^k$ (for $k\geq 1$, here we choose $k=2$)
is Schur-convex \cite{Bengtsson,Bhatia}, we have
\begin{align}
f(\mathbf{e}_0)=\frac{1}{d}\leq f(\mathbf{e}_{\mathcal{I}})=\sum_{\mu=0}^{d-1}|\rho_{\mu\mu}|^2.
\end{align}
Therefore, the maximum value of $C_{l_2}$ for a fixed $\rho$ is given by
\begin{align}
C_{l_2}^{\textrm{max}}(\rho)=\textrm{tr}(\rho^2)-\frac{1}{d},
\end{align}
which is only dependent on the purity of the density matrix and is thus reminiscent of
the results and discussions in Ref. \cite{Singh2015,Yao2016}.

\section{$l_1$ norm of coherence}\label{sec4}
In this section we concentrate on the $l_1$ norm of coherence $C_{l_1}$. Among all the valid quantifiers,
the concept of quantum coherence is more directly embodied in the mathematical definition of $C_{l_1}$, due to the fact that
any nonzero off-diagonal elements of a density matrix will definitely contribute to the ``nonclassicality''
in a given basis. However, despite the simple structure of $C_{l_1}$, it seems difficult to immediately find out
in which basis $C_{l_1}(\rho)$ achieves the maximum value for a given state. Indeed, in view of Theorem \ref{T1},
it is natural to assume the optimal basis is also related to the standard basis by a compound unitary operator,
e.g., $W=VU^\dagger$, where $V$ still diagonalizes the density matrix $\rho$ but the structure of $U$ is unknown.
At this stage, the transformed state is given by
\begin{align}
W^\dagger \rho W=U V^\dagger\rho V U^\dagger=U \Lambda U^\dagger.
\end{align}
Using the Einstein summation convention, the elements of $U \Lambda U^\dagger$ are of the form
\begin{align}
[U \Lambda U^\dagger]_{ij}&=[U]_{ik}\lambda_k\delta_{kl}[U^\dagger]_{lj}=\sum_k\lambda_kU_{ik}U_{jk}^\ast.
\end{align}
Therefore, the $l_1$ norm of coherence is equal to
\begin{align}
C_{l_1}(\rho)=2\sum_{i<j}\left|\sum_k\lambda_kU_{ik}U_{jk}^\ast\right|.
\label{L1}
\end{align}
Nevertheless, so far what we know about $U$ is only the unitary property, that is, $\sum_kU_{ik}U_{jk}^\ast=\delta_{ij}$.

The mathematical subtlety does not prevent us from \textit{guessing} the structure of the unitary matrix $U$.
The first thing coming into our sight is the universal freezing phenomenon that occurs for quantum
correlation or quantum coherence measures \cite{Aaronson2013,Cianciaruso2015,Bromley2015,Yu2016a,Silva2016}.
Here the word \textit{universal} means that under certain initial conditions this phenomenon will inevitably occur
independently of the adopted measures, e.g., it is a common feature of all known \textit{bona fide} measures.
This consistency makes one wonder whether the optimal basis for $C_\mathcal{R}$ also leads to the maximal value
of $C_{l_1}$. However, in the following it will be illustrated that the same basis which is
optimal for $C_R$ is also optimal for $C_{l_1}$ only in the case of qubit and pure states. Moreover, although this basis
corresponds to a stationary point for the optimization problem, there are
numerical evidences that for high-dimensional systems it does not
represent a global maximum for $C_{l_1}$.

\subsection{Qubit and pure states}
For a qubit system, the general one-qubit unitary operator can be parameterized as
\begin{align}
U=e^{\textrm{i}\varphi}
\left(\begin{array}{cc}
a & b \\
b^\ast & -a^\ast
\end{array}\right)
\end{align}
where $|a|^2+|b|^2=1$. Note that the diagonalization process of $\rho$ can be absorbed into the unitary transformation
due to the basis change. The transformed state $U \Lambda U^\dagger$ is given by
\begin{align}
U\Lambda U^\dagger=&
\left(\begin{array}{cc}
a & b \\
b^\ast & -a^\ast
\end{array}\right)
\left(\begin{array}{cc}
\lambda_0 & 0 \\
0 & \lambda_1
\end{array}\right)
\left(\begin{array}{cc}
a^\ast & b \\
b^\ast & -a
\end{array}\right) \nonumber\\
=&\left(\begin{array}{cc}
|a|^2\lambda_0+|b|^2\lambda_1 & ab\lambda_0-ab\lambda_1 \\
a^\ast b^\ast\lambda_0-a^\ast b^\ast\lambda_1 & |b|^2\lambda_0+|a|^2\lambda_1
\end{array}\right)
\end{align}
It follows that $C_{l_1}(\rho)$ for this particular basis (associated with $VU^\dagger$) is
equal to
\begin{align}
\mathcal{O}_2=2|ab||\lambda_0-\lambda_1|\leq|\lambda_0-\lambda_1|.
\end{align}
The above inequality is satisfied as an equality when $|a|=|b|=1/\sqrt{2}$. Indeed, such a unitary matrix is
equivalent to the dephased form, e.g., the Fourier matrix $\mathbf{F}_2$. Therefore, we proved that
this basis is indeed optimal for general qubit states.

Besides, it is easy to see that this basis also holds for the case of pure states in arbitrary dimensions,
since any pure state has only one nonzero eigenvalue, e.g., $\lambda_0=1$ and $\lambda_i=0$ ($1\leq i\leq d-1$).
From Eq. (\ref{L1}), we obtain
\begin{align}
\mathcal{O}_d^{\textsc{pure}}=2\sum_{i<j}\left|F_{i0}F_{j0}^\ast\right|=d-1,
\end{align}
where we denote the elements of the Fourier matrix by $[\mathbf{F}_d]_{ij}=F_{ij}$.
Note that the value of $C_{l_1}$ is upper bounded by $d-1$, and moreover,
the maximal coherence value of $C_{l_1}$ can only be assigned to the maximally coherent states \cite{Peng2016,Napoli2016}.
Thus, this result implies that every pure state in finite dimensions can be represented as a maximally coherent state through the change of basis
and this optimal basis is the one defined by the Fourier matrix (in fact, by all complex Hadamard matrices).

However, the optimization problem starts to become complicated even for general mixed qutrit states. In fact, a complete parameterization
of the space of unitary matrices includes $d^2$ independent real parameters (e.g., Euler angles) \cite{Zyczkowski1994,Dita2003}.
Nevertheless, for reference, here we still present the analytical expression of $O_d$, which is
associated with the specific basis $W=V\mathbf{F}_d^\dagger$ (see Appendix \ref{A1} for details)
\begin{align}
\mathcal{O}_d=\sum_{n=1}^{d-1}\sqrt{\sum_{i=0}^{d-1}\lambda_i^2+\sum_{k\neq l}^{d-1}\lambda_k\lambda_l\cos\left[\frac{2\pi n}{d}(k-l)\right]}.
\end{align}

\subsection{Lagrange multiplier method}
In essence, the issue discussed in this work is an optimization problem. More precisely,
we pursue the maximum value of Eq. (\ref{L1}) subject to the unitarity property $\sum_kU_{ik}U_{jk}^\ast=\delta_{ij}$.
Therefore ,we introduce the Lagrange function
\begin{align}
\mathcal{L}=&\sum_{i\neq j}\left|\sum_k\lambda_kU_{ik}U_{jk}^\ast\right|-\sum_{i,j}\alpha_{ij}\left(\sum_kU_{ik}U_{jk}^\ast-\delta_{ij}\right),\nonumber\\
=&\sum_{i,j}\left|\Theta_{ij}\right|-1-\sum_{i,j}\alpha_{ij}\left(\sum_kU_{ik}U_{jk}^\ast-\delta_{ij}\right)
\end{align}
where $\alpha_{ij}$ are the Lagrange multipliers and to simplify the notation, we define
\begin{align}
\Theta_{ij}=\sum_k\lambda_kU_{ik}U_{jk}^\ast.
\end{align}
Note that $\Theta_{ij}$ is the matrix element of the given density matrix with respect to
the basis $\{VU^\dagger|i\rangle\}$ and thus the symmetry $\Theta^\ast_{ij}=\Theta_{ji}$ holds.
The (local) extreme value of $C_{l_1}$ corresponds to a stationary point for the Lagrange function $\mathcal{L}$.
The first-order partial derivatives are given by
\begin{align}
\frac{\partial\mathcal{L}}{\partial U_{mn}}=\sum_j\frac{\lambda_nU^\ast_{jn}\Theta_{jm}}{|\Theta_{jm}|}-\sum_j\alpha_{mj}U^\ast_{jn}=0,\label{condition1} \\
\frac{\partial\mathcal{L}}{\partial U^\ast_{mn}}=\sum_j\frac{\lambda_nU_{jn}\Theta_{mj}}{|\Theta_{mj}|}-\sum_j\alpha_{jm}U_{jn}=0.\label{condition2}
\end{align}
Multiplying Eq. (\ref{condition1}) by $U_{kn}$ and summing over $n$, we obtain
\begin{align}
\alpha_{mk}=\sum_j\frac{\Theta_{jm}\Theta_{kj}}{|\Theta_{jm}|}.
\end{align}
Similarly, from Eq. (\ref{condition2}) we have
\begin{align}
\alpha_{km}=\sum_j\frac{\Theta_{jk}\Theta_{mj}}{|\Theta_{mj}|}.
\end{align}
Therefore, the very condition for the local extremum can be cast as
\begin{align}
\sum_j\frac{\Theta_{jm}\Theta_{kj}}{|\Theta_{jm}|}=\sum_j\frac{\Theta_{kj}\Theta_{jm}}{|\Theta_{kj}|}.
\label{summation}
\end{align}

Now we can demonstrate that this condition is indeed fulfilled by the Fourier matrix.
For $\mathbf{F}_d$, the matrix elements $\Theta_{ij}$ reduce to
\begin{align}
\Theta_{ij}=\frac{1}{d}\sum_{k=0}^{d-1}\lambda_k\omega^{(i-j)k}.
\end{align}
Therefore, apart from $\Theta_{ij}=\Theta^\ast_{ji}$, there are two additional properties possessed by $\Theta_{ij}$:
(i) \textit{periodic property} $\Theta_{i+d,j}=\Theta_{i,j+d}=\Theta_{ij}$; (ii) \textit{circulant property}
$\Theta_{i,j}=\Theta[(i-j)\,\textrm{mod}\,d]$, which means that the value of $\Theta_{ij}$ is only dependent on the difference
of subscripts.
Due to the periodic property (i), the summation term in Eq. (\ref{summation}) is also a periodic function.
Thus the summation over $j$ can be rearranged to any such region $\{r,r+1,\ldots,r+d-1\}$ for arbitrary integer $r$. By defining $r=k+m-d+1$,
the left hand side of Eq. (\ref{summation}) amounts to
\begin{align}
\sum_{j=0}^{d-1}\frac{\Theta_{jm}\Theta_{kj}}{|\Theta_{jm}|}=\sum_{j=m+k-d+1}^{m+k}\frac{\Theta_{jm}\Theta_{kj}}{|\Theta_{jm}|}
=\sum_{j=0}^{d-1}\frac{\Theta_{m+k-j,m}\Theta_{k,m+k-j}}{|\Theta_{m+k-j,m}|},
\end{align}
where in the last equality we have made the substitution $j\rightarrow m+k-j$. Finally, the circulant property (ii)
guarantees that Eq. (\ref{summation}) indeed holds for the Fourier matrix.

Moreover, we observe that the elements $\Theta_{ij}$ constitute a celebrated \textit{circulant matrix},
which is a special kind of Toeplitz matrix \cite{Gray} (see Appendix \ref{A2}).  It is noteworthy that
circulant matrices have many significant connections to problems in physics, image processing, cryptography and geometry.
For more details and further discussions, we refer the readers to the book by P. J. Davis \cite{Davis}.
In summary, the above results can be recapitulated into the following theorem.

\begin{theorem}
The basis associated with the unitary matrix $W=V\mathbf{F}^\dagger_d$ is (at least) a stationary point
for the $l_1$ norm of coherence. Moreover, the transformed state $W^\dagger\rho W$ is a circulant matrix in the standard basis.
\label{T2}
\end{theorem}

\section{Numerical simulations}\label{sec5}
To verify whether or not this particular basis is global optimal, we can perform a numerical simulation aiming at exhausting the different fiducial bases.
As argued in Section \ref{sec2}, the choice of a random basis is equivalent to uniformly sampling
an element from the group of unitary matrices. To generate random unitary matrices, here we adopt
a simple method proposed by Mezzadri, which is constructed according to the \textit{Haar measure} \cite{Mezzadri2007}.
Such a space of unitary matrices is usually referred to as the circular unitary ensemble (CUE) \cite{Mehta2004}.

For a general qutrit state $\rho$, the analytical expression of
$\mathcal{O}_{3}$ is given by
\begin{align}
\mathcal{O}_3(\rho)=
\sqrt{2}\sqrt{(\lambda_0-\lambda_1)^2+(\lambda_0-\lambda_2)^2+(\lambda_1-\lambda_2)^2},
\end{align}
where $\lambda_0$, $\lambda_1$ and $\lambda_2$ are the (fixed) eigenvalues of $\rho$.
As a typical example, the vector of eigenvalues is chosen to be $\boldsymbol{\lambda}=(0.5,0.3,0.2)$
in descending order. We have run the simulation program from $10^2$ to $10^8$ times (see Table \ref{tabular1}).
Two observations caught our attention: (i) with the increasing count of randomly generated unitary matrices,
the maximal values found in the simulations are getting closer to \textit{but never exceed }
the corresponding value of $\mathcal{O}_3$; (ii) the corresponding unitary matrices are also becoming closer to
the complex Hadamard matrices. For instance, among $10^8$ random unitary matrices, the optimal one takes the form
\begin{align}
|U|=
\left(\begin{matrix}
0.578963 & 0.578962 & 0.574112 \\
0.575579 & 0.576753 & 0.579711 \\
0.577504 & 0.576332 & 0.578213 \\
\end{matrix}\right),
\end{align}
where $|U|$ is the matrix of entrywise absolute values of $U$ and note that $1/\sqrt{3}\approx0.57735$. It is
worth emphasizing that for $d=2, 3, 5$, all complex Hadamard matrices are isomorphism to the Fourier matrix,
which implies it represents the only equivalence class of complex Hadamard matrices \cite{Haagerup1997}.
Moreover, the expression of $\mathcal{O}_3$ remains unchanged with respect to permutations of the
vector of eigenvalues $\boldsymbol{\lambda}=(\lambda_0,\lambda_1,\lambda_2)$. There facts present strong evidences
that for $d=3$ the Fourier matrix is optimal for the $l_1$ norm of coherence.
\begin{table}[ht]
\caption{Numerical results for $d=3$, where $\boldsymbol{\lambda}=(0.5,0.3,0.2)$.
The corresponding value of $\mathcal{O}_3$ is $\sqrt{2(0.09+0.04+0.01)} \approx 0.52915$.}
\begin{tabular}{p{2cm}<{\centering}|p{6cm}<{\centering}}
  \hline \hline
  Counts & The found maximal values \\ \hline
  $10^2$ & 0.508397 \\
  $10^3$ & 0.528649 \\
  $10^4$ & 0.528674 \\
  $10^5$ & 0.528876 \\
  $10^6$ & 0.529063 \\
  $10^7$ & 0.529139 \\
  $10^8$ & 0.529144 \\
  \hline \hline
\end{tabular}
\label{tabular1}
\end{table}

\begin{figure}[htbp]
\begin{center}
\includegraphics[width=0.40\textwidth]{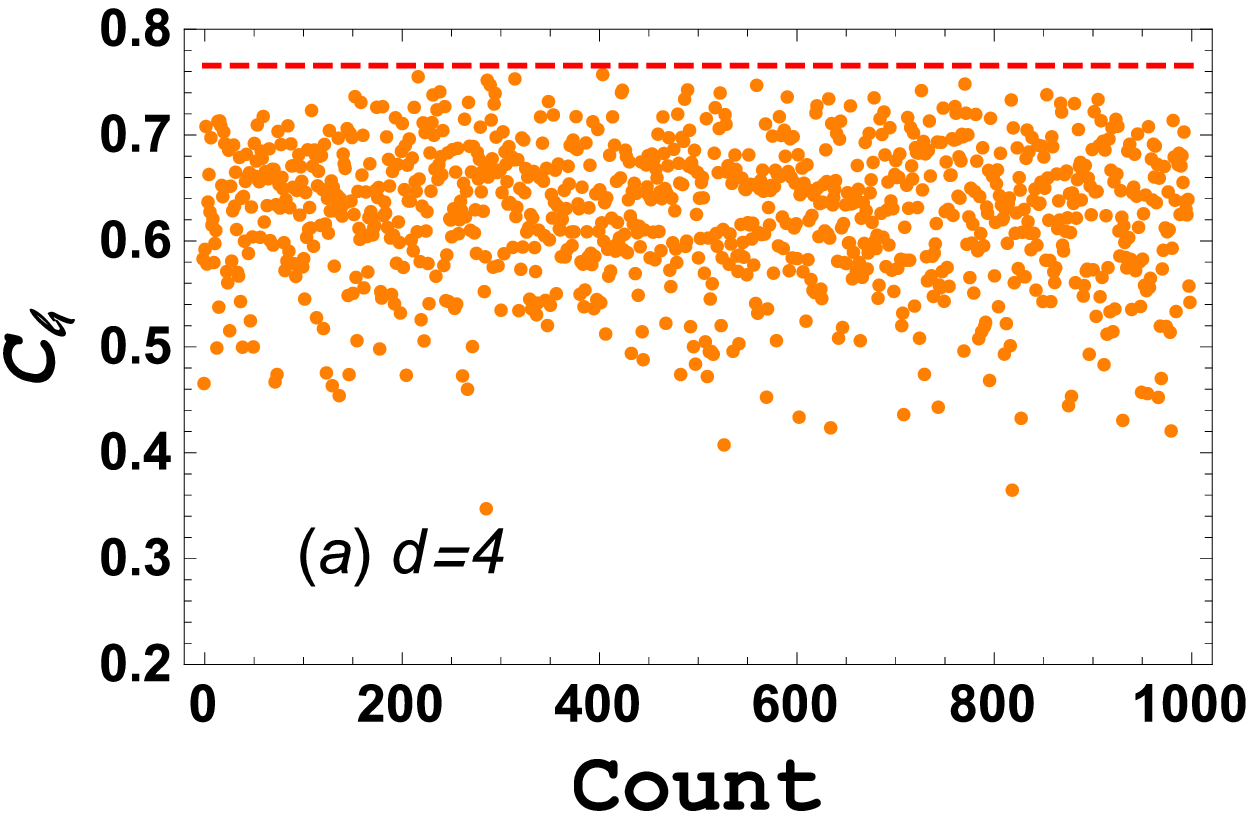}\\
\includegraphics[width=0.40\textwidth]{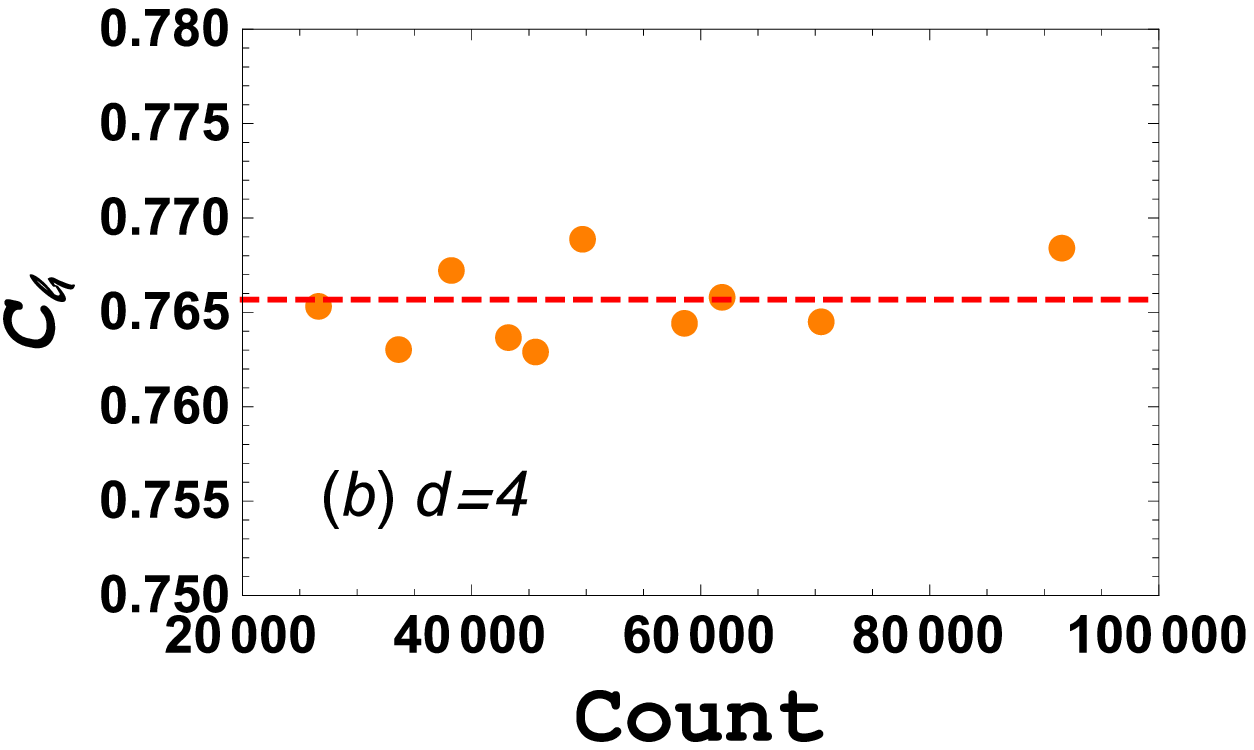}
\end{center}
\caption{(Color online) The horizontal coordinate represents random runs. (a) No violation is observed
for $10^3$ (in fact $10^4$) runs; (b) we picked up the top ten values of $C_{l_1}$ for $10^5$ runs and
totally four violations are observed. Here $\boldsymbol{\lambda}=(0.4,0.3,0.2,0.1)$ and the dashed red line
denotes the corresponding value of $\mathcal{O}_4\approx0.765685$.
}\label{fig1}
\end{figure}

However, for $d=4$ the situation is totally different from the former case. For later discussion, here we present the analytical formula
of $\mathcal{O}_4$ as
\begin{align}
\mathcal{O}_4=2\sqrt{(\lambda_0-\lambda_2)^2+(\lambda_1-\lambda_3)^2}+|\lambda_0-\lambda_1+\lambda_2-\lambda_3|.
\end{align}
First, as for the example chosen in Figure \ref{fig1}, we have not observed any violation of $\mathcal{O}_4$ up to $10^4$ random runs
(here we only plotted $10^3$ runs for simplicity and the statistics of $10^4$ is similar). Yet from $10^5$ runs we begin to
observe the violations of $\mathcal{O}_4$ (see Figure \ref{fig1}). Up to $10^6$ runs, totally $40$ violations can be observed.
Corresponding to the maximal value of $C_{l_1}$ ($\approx0.771506$) found in our simulation,
the matrix of entrywise absolute values of $U$ is of the form
\begin{align}
|U|=
\left(\begin{matrix}
0.374814 & 0.722579 & 0.0537192 & 0.578367 \\
0.588384 & 0.047510 & 0.690752  & 0.417623 \\
0.400215 & 0.667041 & 0.190933  & 0.598689 \\
0.594261 & 0.175156 & 0.695357  & 0.364216 \\
\end{matrix}\right),
\end{align}
which indicates that $U$ is far from being a complex Hadamard matrix.

Moreover, when considering the whole class of complex Hadamard matrices that is equivalent to the Fourier matrix (e.g., $H=D_1P_1\mathbf{F}_dP_2D_2$),
only the permutation $P_2$ contributes to the the coherence value of $H\Lambda H^\dagger$, since $D_1$ and $P_1$ are incoherent unitary operations \cite{Peng2016}.
In fact, the unitary matrix $V$ can also provide this freedom by rearranging the order of its columns.
Therefore, the larger value in the equivalence class would be given by
\begin{align}
\widetilde{\mathcal{O}_d}=\max_{\pi\in\mathcal{P}}\mathcal{O}_d[\pi(\boldsymbol{\lambda})],
\label{update}
\end{align}
where $\mathcal{P}$ is the set of all permutations of the vector $\boldsymbol{\lambda}$.
However, by convention we perviously assume that the vector $\boldsymbol{\lambda}$ is in descending order and
for the example raised in Figure \ref{fig1} this order already gives the maximum value in Eq. (\ref{update}).
Thus, the numerical simulations demonstrate that in general not only the Fourier matrix but also the whole equivalent class
of complex Hadamard matrices are not optimal for the $l_1$ norm of coherence.

Besides, we also performed a simulation for $d=5$ and observed that only up to $10^4$ runs
the maximal value found in the simulation already violates that of $\mathcal{O}_d$
(for a chosen eigenvalue-vector $\boldsymbol{\lambda}=(0.30,0.25,0.20,0.15,0.10)$).
For higher dimensions ($d\geq6$), the simulations can also be carried out via our program, but it is a very time-consuming task.
For given density matrices in $d\geq6$, it is still easy to find the violations of $O_d$,
which implies the Fourier matrix (or the Fourier family) is not global optimal.
However, for example, there exist at least six known distinct equivalence classes for $d=6$ \cite{Bengtsson2007}.
Thus, for high dimensions ($d\geq6$), the above evidences do not exclude the possibility that other \textit{inequivalent} class
of complex Hadamard matrices may result in the global optimal coherence value.
Note that the full construction and classification of complex Hadamard matrices in arbitrary finite dimensions
is still an open question and this problem is unsolved even for $d=6$ \cite{Bengtsson2007,Szollosi2011}.
Therefore, for high dimensions, a much more complicated numerical method is needed to verify
the optimality of some other classes of complex Hadamard matrices.

\section{CONCLUSIONS}\label{sec6}
In this work, we concentrate on the following question: for a given density matrix, in which basis the valid measures of quantum coherence
will achieve the maximum values? On one hand, we have proved that all the bases associated with the unitary operator $VH^\dagger$
is genuinely optimal for the relative entropy of coherence and $l_2$ norm of coherence, where $H$ represents all
complex Hadamard matrices. Indeed, this result stems from the fact that the columns (or rows) of complex Hadamard matrices
are mutually unbiased with the standard basis. On the other hand, although we also proved that this type of bases are
still optimal for general qubit states and pure states in arbitrary dimensions, numerical simulations show that
the Fourier matrix (and its equivalent class) is only suboptimal for our purpose, especially in high-dimensional Hilbert space.
In contrast to the freezing phenomenon for all coherence measures \cite{Bromley2015,Yu2016a,Silva2016}, this result is somewhat counter-intuitive and
indicates that the condition for achieving maximum values of coherence measures is not universal but measure-dependent.
Quite recently, Zanardi \textit{et al.} investigated the coherence power of quantum unitary operators and
they found that all complex Hadamard matrices have maximal coherence generating power \cite{Zanardi2016}.
However, the quantity defined in [43] involves an ensemble
averaging process over all pure states, while in this work we consider
the coherence generating power of a unitary operation with respect to an arbitrary fixed state.

In view of these results, it is worth pointing out that there exist several interesting connections between our work and
some previous findings.
First, since the issue discussed in this work can also be regarded as a coherence-creating
problem, the method raised in Ref. \cite{Misra2016} is a particular case of Theorem \ref{T1}, by noting that
the basis $\{|\phi_j\rangle\}$ used to construct the optimal unitary operation is just
induced by the Fourier matrix
\begin{align}
|\phi_j\rangle=\mathbf{F}_d|j\rangle=\frac{1}{\sqrt{d}}\sum_{i=0}^{d-1}\omega^{ij}|i\rangle=
\mathbb{Z}_d^j|\phi_0\rangle,
\end{align}
where $\mathbb{Z}_d$ is the generalized Pauli operator and satisfies $\mathbb{Z}_d|j\rangle=\omega^j|j\rangle$.
In some other context, the basis introduced in Theorem \ref{T1} is also termed as the \textit{contradiagonal} basis \cite{Lakshminarayan2014}.
Second, the maximum achievable coherence values pursued in this work can also be viewed as a \textit{basis-independent}
quantity. This is thus reminiscent of the concepts introduced in Ref. \cite{Singh2015,Yao2016}, that is
\begin{align}
C_{P}(\rho)=&\sqrt{(d-1)(d\textrm{tr}\rho^2-1)},\\
C_{F}(\rho)=&\sqrt{\frac{d}{d-1}}\left\|\rho-\frac{\openone}{d}\right\|_2,
\end{align}
where $C_{P}(\rho)$ denotes the upperbound of the $l_1$ norm of coherence for fixed mixedness in a system while $C_{F}(\rho)$
characterizes to what extent the given state deviates from the maximally mixed state. In fact, there exists a simple direct relationship
between them $C_{P}(\rho)=(d-1)C_{F}(\rho)$, which indicates that the basis-independent quantity $C_{F}(\rho)$
can also be viewed as a \textit{renormalized} measure of the maximal coherence contained in a given state.
Intriguingly, we found that for $d=2$ and $3$, the formula of $C_{P}(\rho)$
coincides with that of $\mathcal{O}_d$, due to an equivalent expression of $C_{F}(\rho)$ \cite{Yao2016}
\begin{align}
C_F(\rho)=\sqrt{\frac{1}{2(d-1)}\sum_{j,k=0}^{d-1}(\lambda_j-\lambda_k)^2}.
\end{align}
Combining with the numerical results, this fact probably implies that the Fourier matrix is optimal
for arbitrary qutrit states. However, an analytical proof is still missing.

Finally, since $\mathcal{O}_d$ is only suboptimal for $l_1$ norm of coherence, the global optimal basis
and the exact structure of the associated unitary matrix is still left as an open question.
We wonder whether this optimal basis can be directly derived from the criteria that any valid coherence
measures should satisfy \cite{Baumgratz2014}.

\begin{acknowledgments}
We gratefully acknowledge the anonymous referee for helpful comments and constructive suggestions.
This research is supported by the National Natural Science Foundation of China (Grant No. 11605166)
and Science Challenge Project (Grant No. JCKY2016212A503).
C.P. Sun also acknowledges financial support from the National 973 program (Grant No. 2014CB921403),
the National Key Research and Development Program (Grant No. 2016YFA0301201),
and the National Natural Science Foundation of China (Grants No. 11421063 and 11534002).
\end{acknowledgments}
%%%%%%%%%%%%%%%%%%%%%%%%%%%%%%%%%%%%%%%%%%%%%%%%%%%%%%%%%%%%%%%%%%%%%%%%%%%%%%%%%%%%%%%%%%%%%
\appendix

\section{Derivation of $\mathcal{O}_d$}\label{A1}
As shown in Theorem \ref{T2}, for the unitary matrix $W=V\mathbf{F}^\dagger_d$ the transformed state $W^\dagger\rho W$
is actually a circulant matrix. In this specific basis, the $l_1$ norm of coherence can be expressed as
\begin{align}
C_{l_1}(\rho)=d\sum_{n=1}^{d-1}|\Theta_{n}|,
\end{align}
where we define $\Theta_{n}=\frac{1}{d}\sum_k\lambda_k\omega^{nk}$. Furthermore, we have
\begin{align}
|\Theta_{n}|=&\frac{1}{d}\sqrt{\left(\sum_{k=0}^{d-1}\lambda_k\omega^{nk}\right)\left(\sum_{l=0}^{d-1}\lambda_l\omega^{-nl}\right)}\nonumber\\
=&\frac{1}{d}\sqrt{\sum_{i=0}^{d-1}\lambda_i^2+\sum_{k\neq l}^{d-1}\lambda_k\lambda_l\omega^{(k-l)n}}\nonumber\\
=&\frac{1}{d}\sum_{n=1}^{d-1}\sqrt{\sum_{i=0}^{d-1}\lambda_i^2+\sum_{k\neq l}^{d-1}\lambda_k\lambda_l\cos\left[\frac{2\pi n}{d}(k-l)\right]},
\end{align}
which is the desired formula.

\section{Circulant matrix}\label{A2}
As a special kind of Toeplitz matrix, the rows (or columns) of a circulant matrix are composed of cyclically shifted versions of a length-$d$ vector.
Namely, a $d$-dimensional circulant matrix $\mathcal{C}$ takes the form
\begin{align}
\mathcal{C}=
\left(\begin{matrix}
c_0 & c_{d-1} & \ldots & c_2 & c_1  \\
c_1 & c_0 & c_{d-1} & \ddots & c_2  \\
\vdots & c_1 & c_0 & \ddots & \vdots  \\
c_{d-2} & \ddots & \ddots & \ddots & c_{d-1}  \\
c_{d-1} & c_{d-2} & \ldots & c_1 & c_0   \\
\end{matrix}\right).
\end{align}
That is, a circulant matrix is fully specified by a vector $\mathbf{c}=\{c_i\}$ and the entries of $\mathcal{C}$
only rely on the difference of the subscript $(i,j)$
\begin{align}
[\mathcal{C}]_{ij}=c_{(i-j)\,\textrm{mod}\,d},
\end{align}
In our context, the elements of the circulant matrix is given by $[\mathcal{C}]_{ij}=\Theta_{ij}=\frac{1}{d}\sum_k\lambda_k\omega^{(i-j)k}$
and especially the entries on the main diagonal are all equal to $c_0=1/d$.

Another important property of circulant matrices is that they can always be diagonalized by the Fourier matrix \cite{Gray,Davis}.
In this work, the diagonalization is of the form
\begin{align}
\mathcal{C}=\mathbf{F}_d\Lambda \mathbf{F}^\dagger_d,
\end{align}
where $\Lambda=\textrm{diag}\{\lambda_0,\lambda_1,\ldots,\lambda_{d-1}\}$ with $\{\lambda_i\}$ the eigenvalues of $\rho$
(and also of $\mathcal{C}$). In fact, for general circulant matrices, the eigenvalues are given by
\begin{align}
\widetilde{\lambda_j}=c_0+c_{d-1}\omega^j+\ldots+c_1\omega^{j(d-1)}=\sum_{k=0}^{d-1}c_{d-k}\omega^{jk}.
\end{align}
It is easy to check that $\widetilde{\lambda_j}=\lambda_j$ in our case by use of the identity Eq. (\ref{orthogonality}).

%%%%%%%%%%%%%%%%%%%%%%%%%%%%%%%%%%%%%%%%%%%%%%%%%%%%%%%%%%%%%%%%%%%%%%%%%%%%%%%%%%%%%%%%%%%%%

\end{document}